# Intrinsic Reliability improvement in Biaxially Strained SiGe p-MOSFETs

S. Deora[1,3], A. Paul[2], R. Bijesh[1], J. Huang[3], G. Klimeck[2], G. Bersuker[3], P. D. Krisch[3] and R. Jammy[3].

[1] Department of Electrical Engineering, Indian Institute of Bombay, Mumbai 400076, India
[2] School of Electrical and Computer Engineering and Network for Computational Nanotechnology, Purdue University, West Lafayette, Indiana 47907, USA.
[3] Sematech, Austin, TX USA, email: shwetadeora@gmail.com

*Abstract*— In this letter we not only show improvement in the performance but also in the reliability of 30nm thick biaxially strained SiGe (20%Ge) channel on Si p-MOSFETs. Compared to Si channel, strained SiGe channel allows larger hole mobility ($\mu_h$) in the transport direction and alleviates charge flow towards the gate oxide. $\mu_h$ enhancement by 40% in SiGe and 100% in Si-cap SiGe is observed compared to the Si hole universal mobility. A ~40% reduction in NBTI degradation, gate leakage and flicker noise (1/f) is observed which is attributed to a 4% increase in the hole-oxide barrier height ($\varphi$) in SiGe. Similar field acceleration factor ($\Gamma$) for threshold voltage shift ($\Delta V_T$) and increase in noise ($\Delta S_{VG}$) in Si and SiGe suggests identical degradation mechanisms.

## I. INTRODUCTION

As the conventional CMOS scaling approaches its limit, interest in alternate high performance material is intensifying. One of the approaches to overcome the limits of device scaling is to increase the performance by integrating new channel material on Si substrate, while maintaining the compatibility with CMOS technology. Silicon-Germanium (SiGe) possesses several attractive properties which make it the natural contender for extending the performance of Si [1]-[5]. Selective, epitaxially strained SiGe is an attractive channel material due to its high hole mobility and compatibility with conventional CMOS processes [1]. However, placement of high-κ directly on top of SiGe will require low defect density in order to maintain low gate leakage ($J_G$) and better device reliability. Mechanisms leading to a low $J_G$ have been identified [3]. However, it is equally important to understand the mechanisms of reliability degradation in these devices [4], [5].

In this letter, we studied the performance, NBTI and flicker noise (1/f) in SiGe channel p-MOSFETs and compared it with Si devices. Apart from the improvement in the performance, it is also demonstrated that $\Delta V_T$, input referred noise magnitude ($S_{VG}$) and $\Delta S_{VG}$ is lower in SiGe compared to Si. However, both SiGe and Si show identical oxide field ($E_{ox}$) acceleration factor ($\Gamma$) for $\Delta V_T$ and $\Delta S_{VG}$, which demonstrates similar degradation mechanisms. Lower $\Delta V_T$ in SiGe is attributed to the increase in the hole-oxide barrier height ($\varphi$), which reduces the hole tunneling probability. Lower $J_G$ supports lower NBTI degradation in SiGe compared to Si.

## II. DEVICE AND MEASUREMENT DETAILS

Experiments were performed on epitaxial 30nm thick SiGe (~20%Ge) grown on (001) Si surface with <110> channel orientation, and with and without Si cap p-MOSFETs [3]. The substrate doping is ~$10^{18}$cm$^{-3}$. The surface was cleaned with HF chemistry, which was immediately followed by a ~3nm thick HfSiO deposition. The gate dielectric was grown by atomic layer deposition. Subsequent processing includes a conventional gate-first metal gate (MG) process with 950°C source/drain annealing for 10s. Gate material used here is 700nm thick TaN with metal gate work-function ($\Phi_m$) of ~4.6eV. A Si cap of 3nm thickness was used.

NBTI stress was done for 1000s at 125°C at different $E_{ox}$. $\Delta V_T$ was measured using a conventional on-the-fly (OTF) $I_{DLIN}$ technique [7]. Channel length and width are 1µm and 10µm respectively, for NBTI and 1/f measurements.

## III. RESULTS AND DISCUSSION

Figure 1(a) shows the measured $\mu_h$ as a function of inversion layer hole density ($N_S$) for Si, SiGe and Si/SiGe p-MOSFETs along with the universal hole mobility. SiGe shows ~1.5X improvement in $\mu_h$ over Si which gets better with Si-cap (~2X) due to the better oxide-semiconductor interface quality. A similar improvement in the drain current ($I_D$) and the transconductance ($g_m$) is also observed (Fig.1 (b)). It is important to note that in spite of poor interfacial quality between SiGe and gate dielectric [3], a good amount of improvement in the performance is observed in the SiGe devices. Tight-Binding (TB) [11], [12] simulations demonstrate ~2.8X reduction in hole transport mass (m*$_{h<110>}$) and a ~36meV strain field splitting of heavy hole (HH) and light hole (LH) in biaxially strained SiGe (Table 1). Using these m*$_h$, the theoretical estimation of the phonon limited $\mu_h$ enhancement [6] ($\frac{\mu_{SiGe}}{\mu_{Si}} = \frac{m_C^{Si} \cdot m_{DOS}^{Si}}{m_C^{SiGe} \cdot m_{DOS}^{SiGe}}$) is ~3.4X (only HH contribution is taken due to the large HH and LH band splitting), where m$_{DOS}$ and m$_C$ are the DOS and conductivity hole mass respectively, for Si and SiGe. A further reduction in $\mu_h$ is expected due to interface scattering (i.e due to N$_{IT}$ or/and the surface roughness scattering (SRS) [17]) which is not taken into account in [6]. Hence, interface scattering limits the improvement of SiGe over Si to only 1.5X as observed experimentally. *Thus, in spite of poor interface at SiGe/high-k as reported in the literature, performance (I$_D$, g$_m$ and $\mu_h$) improves considerably in biaxial compressive strained SiGe p-MOSFETs compared to Si. Si-cap further enhances performance due to the suppression of interface scattering.*

NBTI study was performed on Si, SiGe and Si/SiGe pMOSFETs. It is now well known that NBTI is oxide field driven (Eox) and by stress bias (V$_G$) [8], [9], [10]. For comparison of NBTI degradation in these devices, effective gate oxide field was estimated from capacitance-voltage (C-V) measurements (Fig.1 (c)). Eox determination was done using a 1D self-consistent Schrodinger-Poisson based C-V simulation. The effective hole mass (m*$_h$) and valence band splitting ($\Delta$Ev) for SiGe devices used for the C-V simulation are obtained from an atomistic 20 band sp$^3$d$^5$s* Tight-Binding based Virtual Crystal Approximation (TB-VCA) model [11], [12]. The m*$_h$ and $\Delta$Ev used for the CV simulations are reported in Table 1. After matching the experimental C-V (Fig.1 (c)), inversion (Q$_{inv}$) and depletion (Q$_{dep}$) charge were determined from the self-consistent calculation, which in turn estimates Eox ~ (Q$_{dep}$+Q$_{inv}$/3)/$\varepsilon_{si}$. We clearly see that C-V of conventional Si device is shifted towards the left by a value ~36mV.

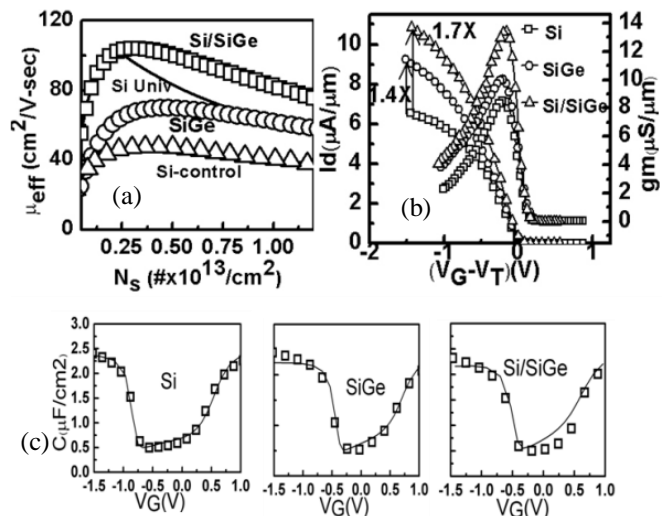

**Figure 1.** (a) Hole mobility (b) Id-Vg and g$_m$ and (c) the capacitance voltage (C-V) for Si, SiGe and Si/SiGe (below).

**Table 1**: Effective hole masses and valence band splitting

| Material | Bands | m*$_h$/m$_o$ | m*$_{DOS}$/m$_o$ | §$\Delta$E$_v$ (eV) |
|---|---|---|---|---|
| Si | HH | 0.276 | 0.653 | 0 |
| Si | LH | 0.214 | 0.24 | 0 |
| Si$_{80}$Ge$_{20}$ | HH | 0.248 | 0.211 | 0.168 |
| Si$_{80}$Ge$_{20}$ | LH | 0.227 | 0.215 | 0.132 |

§These band offsets are w.r.t the unstrained bulk Si VB value of 0eV

Figure 2(a) shows $\Delta$V$_T$ as a function of stress time for Si and SiGe devices for different V$_G$ stress at constant temperature. Irrespective of the channel material, $\Delta$V$_T$ follows power law time dependence with power law time exponent (*n*) ~0.155 - 0.16 (matching that of the reaction-diffusion (R-D) model) for different V$_G$ stress values [9], [10], [13]. It is to be noted that extracted *n* is independent of stress V$_G$, which suggests absence of bulk trap generation [16] in the range of stress bias used for this study. Figure 2(b) shows the Eox dependence of $\Delta$V$_T$ for all the devices. The E$_{OX}$ dependent slope ($\Gamma$) remains the same, irrespective of the channel material. However, a ~4X reduction in $\Delta$V$_T$ is observed for SiGe compared to Si. NBTI involves defect creation by the dissociation of Si-H bonds and subsequent transport of H [9], [13]. The difference in interface defect ($\Delta$N$_{IT}$) between SiGe and Si is mainly expected to arise from the difference in Si-H bond dissociation rate (k$_F$) since the diffusion medium (oxide) is the same. The NBTI degradation rate can be approximated as, $k_f \sim (N_0 * N_s * P_T)$ [13], where N$_0$= Si-H bond number, N$_S$ =

inversion hole density and $P_T$ ($\sim \exp(-(m_{ox}\varphi)^{0.5})$) is the field independent prefactor for hole tunneling probability, $m_{ox}$ = hole effective mass in oxide and $\varphi$ = hole barrier height. Since $N_S$ (obtained from C-V) is found to be constant at a given Eox for all channel materials, $\Delta V_T$ depends only on $N_0$ and $P_T$. The gate leakage current ($J_G$) is a direct indicator of $P_T$. $J_G$ reduces by ~4X in SiGe compared to Si (Fig.3) which is attributed to two factors **(a)** increase in $\varphi$ (0.156eV increase) and **(b)** reduction in hole tunneling mass $m^*_{<001>}$ (1.12X reduction). Factor (a) has a stronger effect (exponential dependence) which suggests that reduction in $P_T$ due to the increase in $\varphi$ is the dominant factor in the reduction of $\Delta V_T$. However, due to the direct placement of high-k on SiGe, $N_0$ is expected to increase (higher $N_{IT}$) and hence increase $\Delta V_T$. But reduction in $P_T$ compensates for the increase in $N_0$ for the devices used in this study. The presence of Si-cap further improves $\Delta V_T$ by ~1.1X compared to SiGe, indicating a better interface quality. *Hence, SiGe intrinsically improves NBTI, which can be verified by the reduction in $J_G$ by approximately the same amount.*

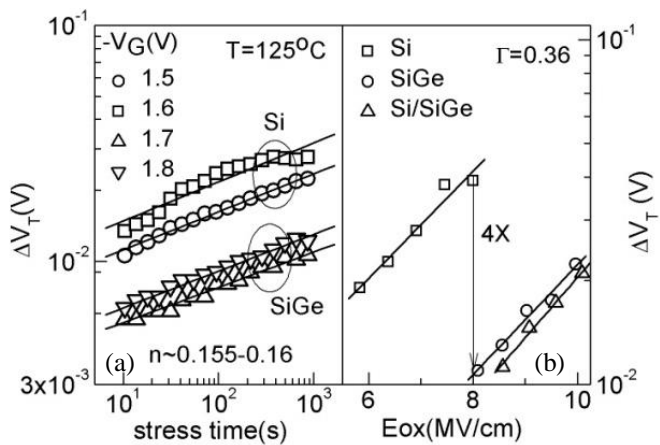

**Figure 2.** (a) $\Delta V_T$ as function of stress time for Si and SiGe. (b) Eox dependence of $\Delta V_T$ for Si, SiGe and Si/SiGe at constant temperature and stress time=1000s.

Robustness of SiGe MOSFETs for analog applications is tested using 1/f noise and impact of NBT stress on $S_{VG}$. Fig. 4(a) shows frequency dependence of $S_{VG}$ at fixed ($V_G$-$V_T$) for all the 3 devices. SiGe and Si/SiGe show a reduction of ~4X and ~10X, respectively in 1/f noise w.r.t Si at a fixed $N_h$ (~$V_G$-$V_T$). From [14], $S_{VG} \propto \frac{1}{\gamma}\int_0^{Vd} N_t^* E_{fs} \frac{1}{N} dV$, $\gamma$ is the attenuation constant of the wave function of the carrier in the oxide, $N_t^*$ is the apparent oxide trap density and $E_{fs}$ is the quasi-Fermi level. First reason for $S_{VG}$ reduction in SiGe is the increase in $\gamma$ ($= \frac{4\pi}{h}\sqrt{2m^*\varphi}$) [14] due to the increase in $\varphi$. Second reason is due to the reduction in $N_t^*$. Contribution to $S_{VG}$ from $N_t^*$ (which is dominant near the $E_{fs}$) gets reduced due to larger separation of the valence band edge from $E_{fs}$ in SiGe devices. Additionally higher $g_m$ for SiGe and Si/SiGe further reduces 1/f noise (since $S_{VG} \sim 1/g_m^2$). Si-cap in Si/SiGe suppresses the direct tunneling, which can further reduce the 1/f noise [15]. Fig. 4(b) shows $\Delta S_{VG}$ as a function of stress Eox after NBT stress of 1000s at 125°C. $\Delta S_{VG}$ shows a ~4X reduction in SiGe w.r.t Si reflecting resilient 1/f noise behavior in SiGe. SiGe and Si devices show identical Eox acceleration factor for $\Delta S_{VG}$, which demonstrate similar degradation mechanism, also corroborated from NBTI study (Fig. 3-b). *Thus, 1/f noise can be reduced by optimizing SiGe layer such that the increase in $N_{IT}$ is compensated.*

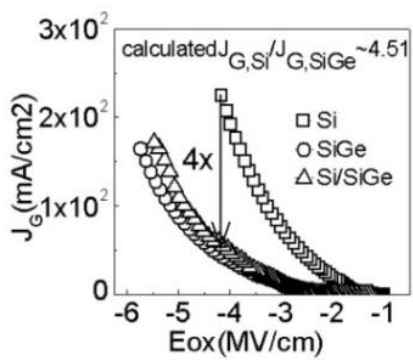

**Figure 3.** Gate leakage current density ($J_G$) as a function of Eox for Si, SiGe and Si/SiGe.

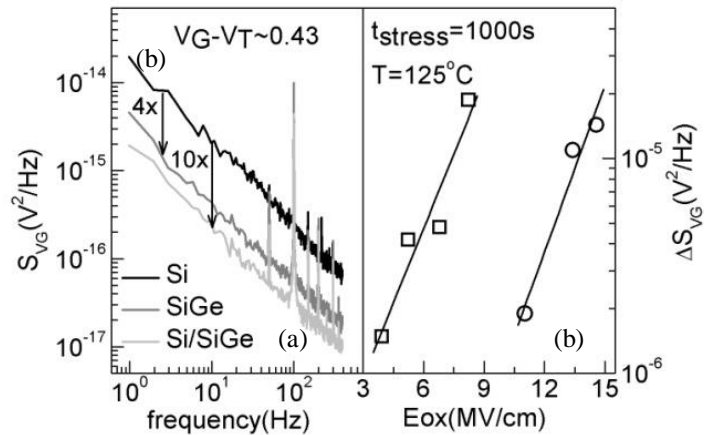

**Figure 4.** (a) $S_{VG}$ as a function of frequency under constant inversion hole density ($N_h = C_{ox}*[V_g-V_T]$). $S_{VG}$ reduces by ~4X and ~10X in SiGe and Si/SiGe respectively compared to Si. (b) $\Delta S_{VG}$ as function of stress $E_{ox}$ measured at constant $|V_G-V_T|$. Similar slopes suggest similar kind of degradation mechanisms for SiGe and Si.

## IV. CONCLUSION

We have shown that biaxially strained SiGe channel shows improvement in performance due to the reduction in the hole transport mass and valance band splitting. There is a significant reduction in $\Delta V_T$ (NBTI), $J_G$, $S_{VG}$ (1/f) and $\Delta S_{VG}$ for SiGe compared to Si. Si-cap on SiGe further improves performance and reliability, but at an added process complexity. Therefore, in spite of high $N_{IT}$ as reported in the literature, intrinsic properties of SiGe (obtained from TB simulations) not only improve the performance but also enhance the device reliability. *By proper device and process optimization [2, 3], SiGe alone can qualify as a promising channel material for future CMOS technology.*

ACKNOWLEDGMENT

The authors would like to acknowledge the reviewers for their useful comments. nanoHUB.org for computational resources. Financial support from GRC, MSD-FCRP, NSF and MIND/NRI are also acknowledged.